\documentclass[12pt]{article}
\usepackage{epsfig}

\def\beq{\begin{equation}}
\def\eeq#1{\label{#1}\end{equation}}
\def\eeqn{\end{equation}}

\def\beqa{\begin{eqnarray}}
\def\eeqa#1{\label{#1}\end{eqnarray}}
\def\eeqan{\end{eqnarray}}

\let\bar=\overbar

\def\etal{{\it et al.}}

\def\Dslash{\not{\hbox{\kern-4pt $D$}}}
\def\dslash{\not{\hbox{\kern-2pt $\del$}}}

\def\ee{e^+e^-}

\def\msb{{\bar{\ssstyle M \kern -1pt S}}}

\def\Title#1{\begin{center} {\Large {\bf #1} } \end{center}}

\begin{document}

\Title{$B \to X_s\ell^+ \ell^-$ at Belle }

\bigskip\bigskip


\begin{raggedright}  

{\it Katsumi Senyo\index{Katsumi Senyo}\\
Department of Physics \\
Nagoya University\\
Nagoya, 464-8602 JAPAN}
\bigskip\bigskip
\end{raggedright}

\section{Introduction}
The electroweak penguin mediated $B \to X_s\ell^+\ell^-$ decay is a
good probe to study beyond the Standard Model physics at low energy.
The effect of new physics could be observed
as corrections to the branching fraction,
di-lepton invariant mass spectrum, and to the Wilson
coefficients such as $C_9$, $C_{10}$, and a phase of $C_7$~[1-10].
Experimentally, the exclusive $B \to K \ell^+ 
\ell^-$ decay was observed by the Belle collaboration for the first
time~\cite{belle:BKll} and has also recently confirmed by the BABAR
collaboration~\cite{babar:fpcp}.  We present here a preliminary
result on a  
search for the inclusive $B \to X_s \ell^+ \ell^-$ decay, which can be
calculated more reliably than exclusive decay modes.

\section{Analysis}
The data sample used in this analysis consists of $43~{\rm fb}^{-1}$ of
$e^+e^-$ collisions at the $\Upsilon(4S)$ resonance collected by
the Belle detector~\cite{nim} at the KEKB storage ring.

The signal is identified as $X_s \ell^+ \ell^-$, where $X_s$ is defined to
be a $K^{\pm}$ or $K_S$ accompanying 0 to 4 pions including one
$\pi^0$ at most, and $\ell^+ \ell^-$ is an oppositely charged electron
or muon pair. 
Charged tracks are required to originate from the
interaction point and be well identified by the Belle particle identification
devices.  Electrons are identified using the ratio between
the energy deposited in the electromagnetic calorimeter and the measured
track momentum($E/p$).  Electrons are required to have momenta above $0.5~{\rm
GeV/}c$.  Muon candidates are
required to reach the iron flux return and have momenta greater than
$1.0~{\rm GeV/}c$.  The muon identification efficiency is about 80 to
85 \% with a fake rate of 1 to 2 \%.  Charged hadrons 
are identified by the combined response of aerogel counters, time
of flight, and $dE/dx$ measured in the drift chamber.  For charged kaons,
we have an efficiency of 85 to 90\% and a pion fake rate that is
less than $10 \%$.
Candidate $\pi^0$'s are identified from pairs of photons with
energy deposits of at least $50~{\rm MeV}$ 
and an invariant mass within $10~{\rm MeV}/c^2$ 
of the nominal $\pi^0$ mass.

Events with leptons from charmonia, such as $J/\psi$ and $\psi '$
are vetoed if the dilepton mass lies around the nominal charmonia
masses.

Background from $q\bar{q}$ continuum events is suppressed
by requiring the ratio of the second to zeroth Fox-Wolfram moments $R_2
\equiv H_2/H_0 < 0.35$ and $|\cos \theta_{thrust}| < 0.85$, where $\theta_{thrust}$ is
the angle between the thrust axis of the $B$ candidate and that of the rest
of the event.  The $|\cos \theta_{thrust}|$ distribution of 
continuum events peaks at 1 while that of signal events is flat.

To reduce backgrounds further and select the best candidate in an event,
we introduce four
kinematic variables: the angles between the $K$ and
$\ell^+$($\theta_{K\ell^+}$), and the $K$ and
$\ell^-$($\theta_{K\ell^-}$), the $B$ flight
direction($\theta_B$), and the energy difference($\Delta E = E_B - E_{beam}$).  
If the $K$ meson and $\ell^+\ell^-$ pair originated from a photon or
$Z$ boson; since these are emitted in opposite directions, the sum
of $\cos \theta_{K\ell^+}$ and $\cos \theta_{K\ell^-}$ should be
negative for the signal. 
The polar angle of $B$ flight direction has a $1 - \cos^2 \theta_B$
distribution while background is flat. 
The signal $\Delta E$ distribution peaks at zero.

A likelihood ratio $LR$ is determined by parameterizing
distributions of three kinematic variables with Monte Carlo simulation
for both the signal and background.  It is expressed as
$ LR = p_{sig}/(p_{sig} + p_{BG})$,
where $p_{sig}$ and $p_{BG}$ are probability density functions for
the signal and background distributions, respectively.  
An $LR$ selection requirement is chosen to
maximize the figure of merit~$S/ \sqrt{S+N}$ in the signal region
where $S$ is the expected signal yield assuming the SM
prediction and $N$ is the number of expected background events.

The invariant mass of the $X_s$ must satisfy $M_{X_s} < 2.1~{\rm
GeV}/c^2$ in order to reject combinatorial background.

After the application of all selection requirements, the reconstruction
efficiencies are 
estimated by Monte Carlo simulation based on the inclusive 
model in the recent paper by Ali~\etal\cite{ali} and based on the
exclusive $B \to K \ell^+\ell^-$ and $B \to K^* \ell^+\ell^-$ model by
Greub~\etal\cite{greub}.  The nominal reconstruction efficiencies are
determined to be $3.6~\%$ for $B \to X_s e^+ e^-$ and $3.8~\%$ for $B
\to X_s \mu^+ \mu^-$.

Most of the background is combinatorial background from $B$
and $D$ decays.  The background is determined by
fitting the distribution of beam constrained mass $M_{bc}$.
The beam constrained mass is expressed as $M_{bc} =
\sqrt{(E_{beam})^2 - 
(p_B)^2}$, where $p_B$ is the B candidate's center-of-mass momentum vector
and $E_{beam}$ is the center-of-mass energy of $B$ meson,
respectively.  In this distribution, the signal peaks at
the $B$ meson mass and is fitted by a Gaussian function while the
combinatorial background is parameterized by a phase space function with a
kinematic threshold~(the ARGUS function).

Background from $B \to X_s \pi^+ \pi^-$ decay can 
make a small contribution to the signal peak in the $M_{bc}$ 
distribution when both the $\pi^+$ and $\pi^-$ are misidentified as
muons.  The muon misidentification rate is determined to be 1 to 2
\% in the laboratory momentum range $p_{\mu} > 1.0 {\rm GeV}/c$.
To estimate the contamination in the signal region, a $B \to X_s
\pi^+ \pi^-$ sample is selected by applying all the signal selection 
requirements except for the muon identification.  The yield is
multiplied by the momentum dependent fake rate for each pion.
This $B \to X_s \pi^+ \pi^-$ background is 
estimated to be 2.4 events and subtracted from the yield of the signal fit.

\section{Results and discussion}
\begin{figure}[b]
\begin{center}
\epsfig{file=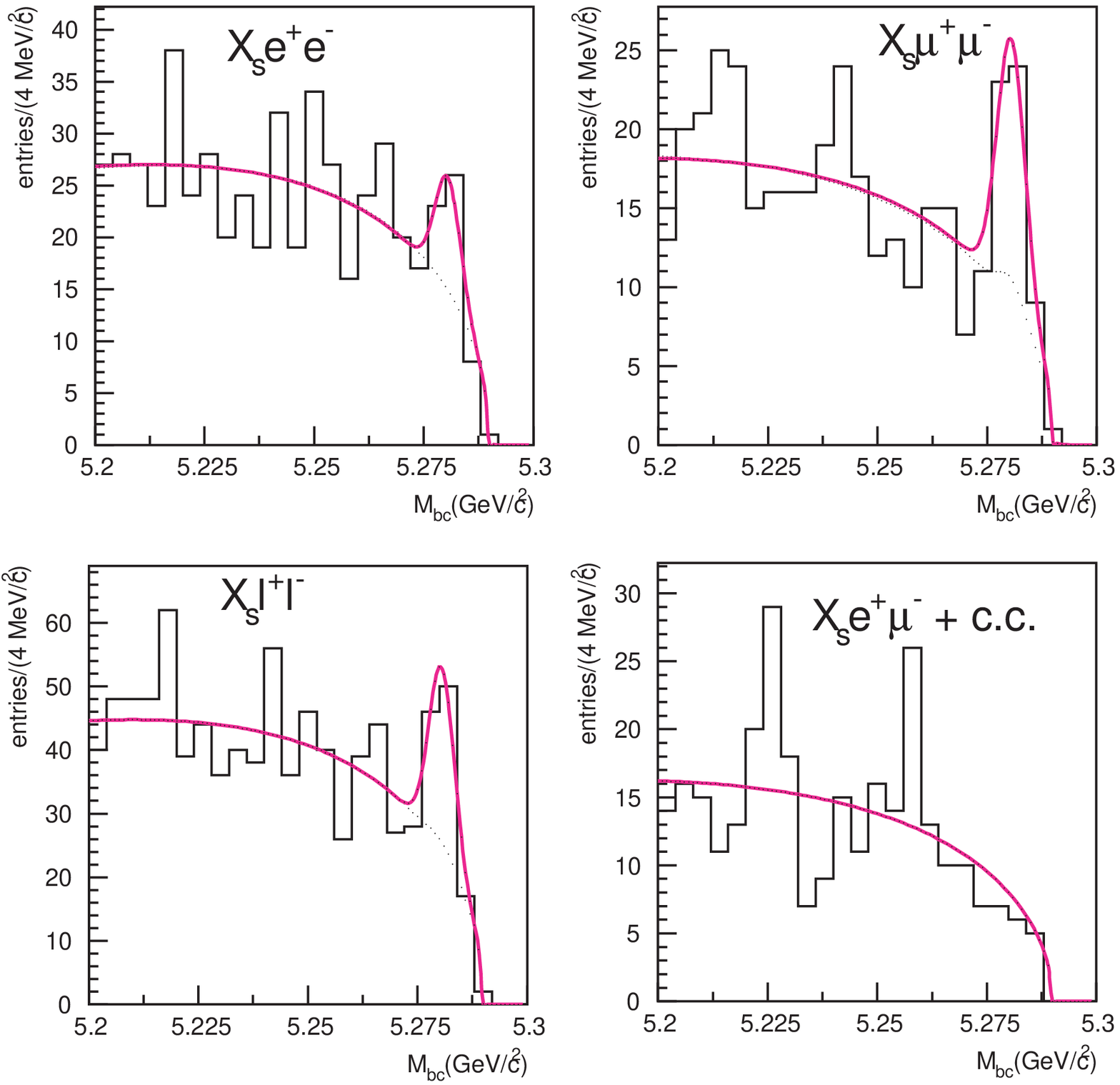,width=10cm}
\caption{$M_{bc}$ distributions and fit results.  Top-left: $X_s e^+ e^-$
 candidates, top-right: $X_s \mu^+ \mu^-$ candidates, bottom-left: $X_s \ell^+ \ell^- = (X_s e^+ e^-) + (X_s \mu^+ \mu^-)$ candidates, and bottom-right: $X_s e^{\pm} \mu^{\mp}$ to estimate combinatorial background.  The significance is determined from the statistical error only.}
\label{fig:results}
\end{center}
\end{figure}

Figure~\ref{fig:results} shows the $M_{bc}$ distribution for $B \to X_s
e^+e^-$, $B \to X_s \mu^+\mu^-$, and for the combined $B \to X_s
\ell^+\ell^-$ samples, respectively.  The
$B \to X_s e^{\pm}\mu^{\mp}$ sample is used to check the background
parameterization.  
The fit results, signal yields, branching fractions, and statistical
significances are shown in
Table~\ref{tab:results}.  

Systematic uncertainties in this measurement
are listed in Table~\ref{tab:systematics}.  The largest contribution to
the systematic uncertainty is the decay modeling for $B \to X_s
\ee$ and $B \to X_s \mu^+ \mu^-$ decays.  

We measure the inclusive $B \to X_s \ell^+ \ell^-$ decay branching
fraction for the first time with a statistical significance of greater
than $4\sigma$.  Preliminary results for branching
fractions are ${\cal B}(B \to X_s \mu^+ \mu^-)=
(8.9^{+2.3+1.6}_{-2.1-1.7}) \times 10^{-6}$ and 
${\cal B}(B \to X_s \ell^+ \ell^-)=
(7.1^{+1.6+1.4}_{-1.6-1.2}) \times 10^{-6}$, where the first error is
statistical and the second is systematic.

The branching fractions and upper limits determined by this study agree
well with theoretical predictions in the SM framework.
We expect that both the theoretical and experimental
errors will decrease so that theories based on
SM can be tested with precision measurements in the near
future. 

\begin{figure}[htcb]
\begin{center}
\begin{tabular}{cc}
\begin{minipage}[b]{0.3\hsize}
\begin{center}
\epsfig{file=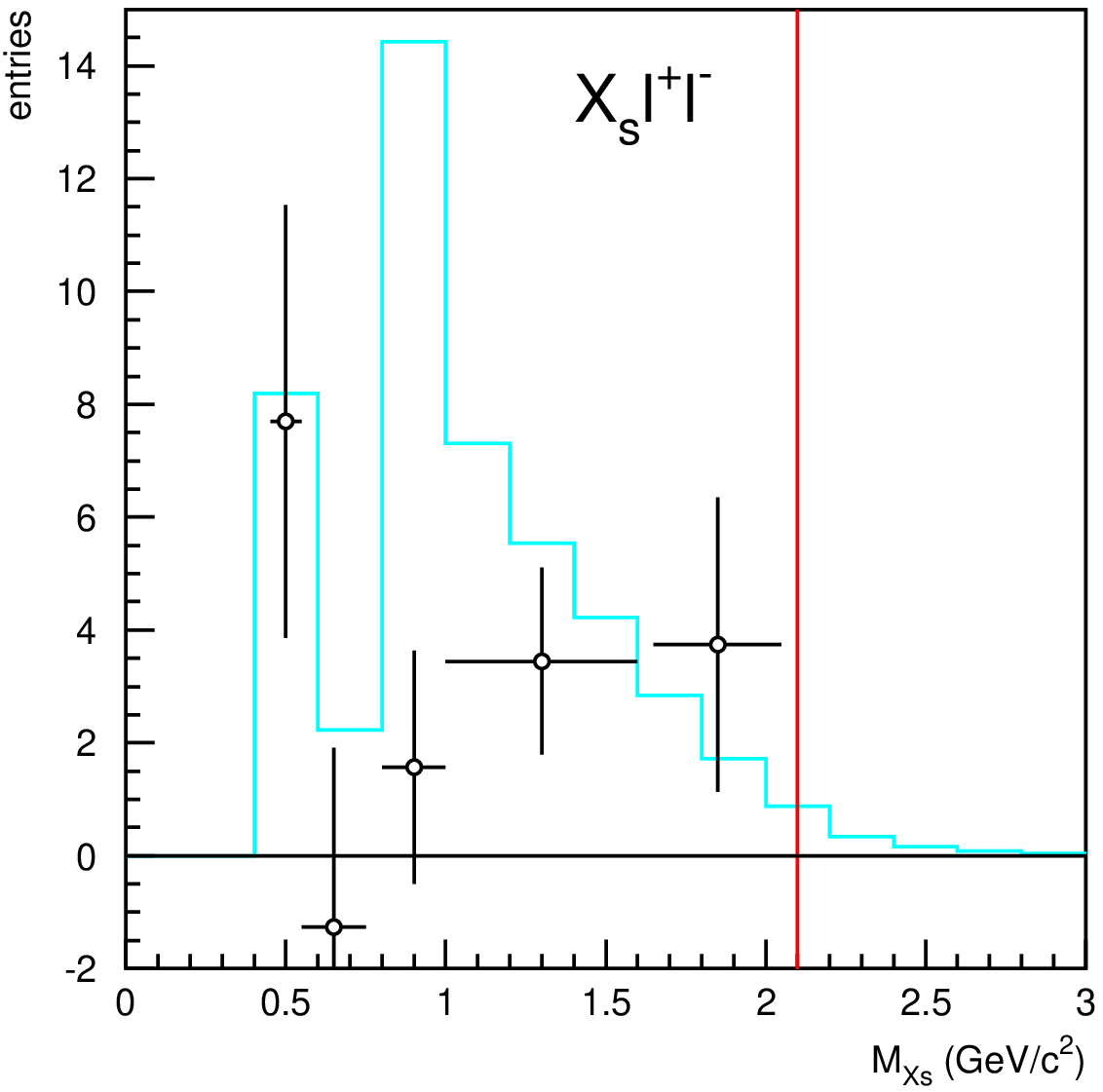,width=\hsize}
\caption{Comparison of the $M_{X_s}$ distribution for data and MC.}
\label{fig:MXs}
\end{center}
\end{minipage}
\enskip
\begin{minipage}[b]{0.3\hsize}
\begin{center}
\epsfig{file=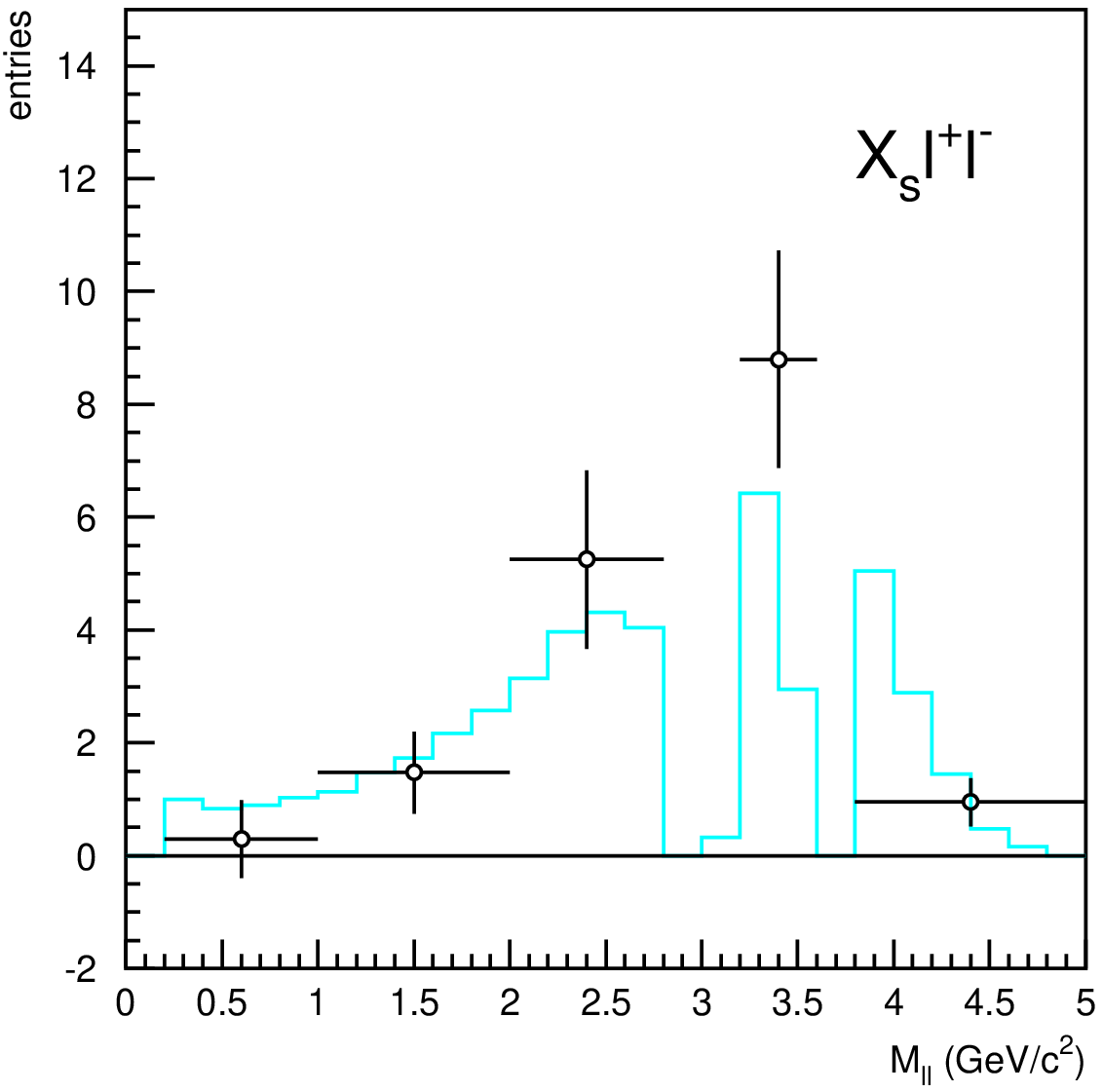,width=\hsize}
\caption{Comparison of the $M_{\ell \ell}$ distribution for data and MC. }
\label{fig:Mll}
\end{center}
\end{minipage}
\end{tabular}
\end{center}
\end{figure}

The $M_{X_s}$ and dilepton invariant mass plots are shown in
Figures~\ref{fig:MXs}~and~\ref{fig:Mll}, respectively.  The
$M_{X_s}$ distribution extends beyond the $K$ and $K^*$ mass region.
The dips around $3.1{\rm GeV}/c^2$ and
$3.7{\rm GeV}/c^2$ in the $M_{\ell \ell}$ distribution are due to $J/\psi$
and $\psi'$ vetoes. 
The experimental and theoretical uncertainties are expected to improve
as more data is accumulated.  Precision measurements can be done for SM
and beyond SM predictions.

In summary, evidence for the inclusive $B \to X_s \ell^+ \ell^-$  decay
has been 
presented for the first time.  The experimental and theoretical
uncertainties are expected to decrease with time.  Precision
measurements can be done in near future that test the Standard Model and
Beyond the Standard Model predictions.


\begin{table}[t]
\begin{minipage}[b]{0.6\hsize}
\begin{center}
\begin{tabular}{llcc}  
\hline
Mode           &  \#signal                  &  B.F.$(\times 10^{-6})$     & Signif. \\ \hline
 $X_s e^+ e^-$   &$16.6^{+8.0+3.9}_{-7.3-3.8}$&  $ < 11.0$          & 2.1    \\
 $X_s \mu^+\mu^-$&$30.7^{+7.9+5.4}_{-7.4-3.8}$&$8.9^{+2.3+1.6}_{-2.1-1.7}$  &   4.4    \\
 $X_s \ell^+\ell^-$&$47.6^{+11.0+9.6}_{-10.4-8.0}$&$7.1^{+1.6+1.4}_{-1.6-1.2}$      & 4.8    \\
\hline
\end{tabular}
\caption{Preliminary results for $B \to X_s e^+ e^-$, $X_s \mu^+ \mu^-$,
 and combined $X_s \ell^+ \ell^-$.  Significance is extracted from
 the statistical error only.  Only a 90\% confidence level upper
 limit is shown for the $B \to X_s e^+e^-$ process.}
\label{tab:results}
\end{center}
\end{minipage}
\enskip
\begin{minipage}[b]{0.4\hsize}
\begin{center}
\begin{tabular}{lcc}  
\hline
Source & $X_s e^+ e^-$ & $X_s \mu^+ \mu^-$ \\
\hline
Tracking & 8.1 \% & 8.0 \% \\
Kaon ID  & 1.9 \% & 2.0 \% \\
Pion ID  & 0.8 \% & 0.8 \% \\
Lepton ID & 3.6 \% & 4.4 \% \\
$K_S$ detection & 2.1 \% & 1.5 \% \\
$\pi^0$ detection & 2.0 \% & 1.6 \% \\
MC stat. & 3.9 \% & 4.1 \% \\
Decay model& $ ^{+14 \%}_{-9 \%}$ &$ ^{+16 \%}_{-12 \%}$ \\ 
\hline
Total & $ ^{+18 \%}_{-14 \%}$ & $ ^{+19 \%}_{-16 \%}$ \\
\hline
\end{tabular}
\caption{Systematic uncertainty summary.}
\label{tab:systematics}
\end{center}
\end{minipage}
\end{table}

\begin{thebibliography}{99}

\bibitem{ali}
A. Ali, E. Lunghi, C. Greub, and G. Hiller, hep-ph/0112300.
 \bibitem{greub}
C. Greub, A. Ioannissian and D. Wyler, Phys. Lett. {\bf B346},149(1995).
\bibitem{Borzumati}
F. Borzumati and C. Greub, Phys. Rev. {\bf D58}, 074004(1998);F. Borzumati and C. Greub, Phys. Rev. {\bf D59}, 057501(1999); M. Ciuchini, G. Degrassi, P. Gambino and G. F. Giudice, Nucl. Phys. {\bf B527}, 21(1998).
\bibitem{Besmer}
T. Besmer, C. Greub and T. Hurth, CERN-TH-2001-136, BUTP-01-12, ZU-TH-15-01, hep-ph/0105292; M. Ciuchini, G. Degrassi, P. Gambino and G. F. Giudice, Nucl. Phys. {\bf B534}, 3(1998), C. Bobeth, M. Misiak and J. Urban, Nucl. Phys. {\bf B567}, 153(2000); F. Borzumati, C. Greub, T. Hurth an D. Wyler, Phys. Rev. {\bf D62}, 075005(2000).
\bibitem{okada}
T.Gotou, Y. Okada and Y. Shimizu, KEK-TH-611, hep-ph/9908499.
\bibitem{ali2}
A. Ali, G. F. Giudice and T. Mannel, Z. Phys. {\bf C67}, 417(1995); A. Ali, T. Mannel and T. Morozumi, Phys. Lett. {\bf B273}, 505(1991); W. Jaus and D. Wyler, Phys. Rev. {\bf D41}, 3405(1990).
 \bibitem{ali3}
A. Ali, P. Ball, L. T. Handoko and G. Hiller, Phys. Rev. {\bf D61}, 074024(2000).
 \bibitem{melikhov}
D. Melikhov and N. Nikitin, Phys. Lett. {\bf B410}, 290(1997).
 \bibitem{ali1997}
A. Ali, G. Hiller, L. T. Handoko and T. Morozumi, Phys. Rev. {\bf D55},
	4105(1997). 
 \bibitem{kruger}
F. Kruger and L. M. Sehgal, Phys. Lett. {\bf B380}, 199(1996).
 \bibitem{belle:BKll}
K. Abe, \etal(Belle Collaboration), Phys. Rev. Lett. {\bf 88}, 021801(2002).
 \bibitem{babar:fpcp}
J. Walsh (BABAR Collaboration), presented at Flavor Physics and CP
Violation(FPCP) at Philadelphia, 2002.

\bibitem{nim}
A. Abashian \etal(Belle Collaboration), Nucl. Instr. and Meth. {\bf
	A479}, 117 (2002).

\end{thebibliography}
\end{document}